%% file: bare_conf.tex
\documentclass[conference]{IEEEtran}
\IEEEoverridecommandlockouts
\usepackage{cite}
\usepackage{amsmath,amssymb,amsfonts}
\usepackage{algorithmic}
\usepackage{graphicx}
\usepackage{textcomp}
\usepackage{xcolor}
\usepackage{subfigure} 
\usepackage{float}
\usepackage[colorinlistoftodos]{todonotes}
\def\BibTeX{{\rm B\kern-.05em{\sc i\kern-.025em b}\kern-.08em
    T\kern-.1667em\lower.7ex\hbox{E}\kern-.125emX}}
\renewcommand{\vec}[1]{\mathbf{#1}}
\begin{document}

\title{A Hybrid Recommender System for Patient-Doctor Matchmaking in Primary Care
\thanks{This work was supported in part by Nova School of Business and Economics through the Data Science for Social Good Europe Fellowship, AWS Cloud Credits for Research by Amazon.com Inc, and Azure Research grant by Microsoft Corporation.}
}

\author{
    \IEEEauthorblockN{Qiwei Han\IEEEauthorrefmark{1}, 
Mengxin Ji\IEEEauthorrefmark{2}, 
    \'{I}\~{n}igo Mart\'{i}nez de Rituerto de Troya\IEEEauthorrefmark{1}, Manas Gaur\IEEEauthorrefmark{3}, Leid Zejnilovic\IEEEauthorrefmark{1}}
    \IEEEauthorblockA{\IEEEauthorrefmark{1}Universidade Nova de Lisboa, School of Business and Economics, \{qiwei.han, itroya, leid.zejnilovic\}@novasbe.pt}
    \IEEEauthorblockA{\IEEEauthorrefmark{2}University of California, Davis, Department of Agricultural and Resource Economics, mji@ucdavis.edu}
  \IEEEauthorblockA{\IEEEauthorrefmark{3}Wright State University, Kno.e.sis Center, manas@knoesis.org}
}
\maketitle

\begin{abstract}
We partner with a leading European healthcare provider and design a mechanism to match patients with family doctors in primary care. We define the matchmaking process for several distinct use cases given different levels of available information about patients. Then, we adopt a hybrid recommender system to present each patient a list of family doctor recommendations. In particular, we model patient trust of family doctors using a large-scale dataset of consultation histories, while accounting for the temporal dynamics of their relationships. Our proposed approach shows higher predictive accuracy than both a heuristic baseline and a collaborative filtering approach, and the proposed trust measure further improves model performance. 
\end{abstract}


\section{Introduction}
\input{1_introduction}

\section{Related Work}
\input{related}

\section{Background and Data}
\input{2_data}

\section{Exploratory Data Analysis}
\input{eda}

\section{Methods}
\input{3_methods}

\section{Results}
\input{4_results}

\section{Conclusions}
\input{conclusions}

%\newpage
\bibliography{bare_conf}
\bibliographystyle{IEEEtran}

\end{document}

%% file: 1_introduction.tex
Primary care serves as patients' first point of contact with the healthcare system and is a continuing focal point of comprehensive, accessible, and community-based care\cite{who}. More than just a gate-keeping process for specialist referrals, it has been widely recognized for its focus on caring for the long-term health of patients rather than solely for treating specific diseases or conditions. As such, primary care helps deliver more equitable health outcomes across populations and meets 80-90\% of individuals' health needs throughout their lives\cite{starfield2005}. To this end, a recent special report from the Economist stated that ``good primary care is an essential precondition for a decent healthcare system'' \cite{economist2018}.

The World Health Organization (WHO) emphasized several defining features for effective and socially productive primary care, including comprehensiveness, person-centeredness, and continuity of care \cite{who2008}. In particular, person-centeredness refers to the ``clinical method of participatory democracy'' that allows patients to participate in decisions that affect their health. For example, patients value more freedom in choosing their primary care physicians\footnote{We use primary care physicians, general practitioners, and family doctors interchangeably throughout this paper.}, with whom they can build an enduring and trusting relationship. This ability to choose not only improves the quality of care and patient satisfaction with primary care physicians and health-care providers \cite{ferrer2005, darin2007}, but also results in increased trust and treatment compliance for better health outcomes \cite{fiscella2004}.

However, it is very challenging for patients to find the right family doctors with whom they can build trusting relationships, particularly when an appropriate matchmaking mechanism is not available. On the one hand, healthcare providers often lack the infrastructure and service design implementations to transform their services to more person-centered approaches, \emph{e.g.} enabling patients to choose their doctor \cite{gottlieb2008}. On the other hand, patients face significant search costs in understanding the competencies of all available doctors and thus resort to word-of-mouth recommendations from friends, relatives, or online reviews to resolve the uncertainty. The barrier between the rapidly changing institutional environment and increasing patient autonomy complicates matchmaking between patients and family doctors.

Moreover, the extent to which matchmaking between patients and family doctors measurably benefits their relationship remains an unresolved issue. Given that trust in patient-doctor relationships plays a central role in improving patients' health outcomes and satisfaction with their care \cite{birkhauer2017}, it would be preferable to match patients with family doctors that they are willing to consult with high trust. However, researchers typically use qualitative analysis based on survey data to examine factors affecting patients' trust in family doctors \cite{croker2013}, while rarely considering rich data such as past patient-doctor interactions that may strongly signal the relationship. For example, repeated interactions between patients and doctors may well represent the continuity of care, which can develop trust over time \cite{tarrant2010}. Also, patients with similar characteristics (\emph{e.g.} gender, age) may exhibit preferences for family doctors with similar characteristics; this is commonly known as ``homophily'' and may influence patients' perception of communication with doctors and quality of care \cite{thornton2011}. Therefore, it is crucial to leverage these healthcare data to contribute to designing matchmaking mechanisms in primary care toward trusting patient-doctor relationships. 

In this paper, we aim to help a leading European healthcare provider transform their primary care health service into a person-centered service by reorganizing their appointment system for general practitioners with the goal of reducing search frictions and becoming more directly accessible. We do so by proposing a hybrid recommender system that automates the matchmaking process between patients and family doctors. In particular, we model patients’ trust in family doctors using a large-scale dataset of consultation histories, while accounting for the temporal dynamics of their relationships. We further combine patient and doctor metadata to model similarities between patients and doctors across social dimensions. This is especially useful when patients have limited prior consultations with family doctors. As such, we can generate personalized doctor recommendations for each patient that they may trust the most. To the best of our knowledge, this work is among the first Data Collaborative initiatives in this European country that exchanges data from a private entity in the health sector to create socially desirable value.

We define several use cases for the matchmaking process, each suited to a different category of patients for whom different degrees of information are available. More specifically, we build a hybrid matrix factorization model to represent patients and doctors as linear combinations of latent embeddings derived from their characteristics and interactions. In combination with a rule-based model selection process, our system provides a unified approach for presenting any patient with a list of personalized doctor recommendations.

Our hybrid approach shows higher predictive accuracy than both a heuristic baseline and a traditional collaborative filtering (CF) recommender system. More interestingly, the proposed quantitative trust measure further improves model performance when incorporated into either hybrid or CF systems. We believe that our approach not only provides a useful application in a critical social service, namely when healthcare providers move towards making their primary care services more patient-centered, but may also shed light on how to leverage large-scale healthcare datasets to quantify critical qualitative factors such as trust in patient-doctor relationships, an area with a paucity of empirical research \cite{pearson2000}.

The rest of the paper is organized as follows. Section 2 reviews the related literature from two distinct areas: the importance of the patient-doctor relationship and applications of recommender systems in the health sector. Section 3 examines the data sources from our partner and the use cases for which we wish to match patients and doctors. Section 4 performs exploratory data analysis to identify critical demographic and transactional characteristics of patients and doctors. Section 5 discusses the methods employed in the hybrid recommender system formulation as well as the novel quantitative measure of trust. Section 6 compares the performance of the proposed models. Finally, Section 7 concludes and discusses the future work.

% % no \IEEEPARstart
% This demo file is intended to serve as a ``starter file''
% for IEEE conference papers produced under \LaTeX\ using
% IEEEtran.cls version 1.7 and later.

% All manuscripts must be in English. These guidelines include complete descriptions of the fonts, spacing, and related information for producing your proceedings manuscripts. Please follow them and if you have any questions, direct them to the production editor in charge of your proceedings at Conference Publishing Services (CPS): Phone +1 (714) 821-8380 or Fax +1 (714) 761-1784.
% % You must have at least 2 lines in the paragraph with the drop letter
% % (should never be an issue)

% \subsection{Subsection Heading Here}
% Subsection text here.

% \subsubsection{Subsubsection Heading Here}
% Subsubsection text here.

%% file: related.tex
\subsection{Patient-Doctor Relationship}
The patient-doctor relationship is a keystone of healthcare \cite{goold1999}. Each time a patient consults with their doctor, they engage in a relationship that directly determines their quality of care and, eventually, their satisfaction with the treatment. Two main factors have been found to contribute to the process by which patient-doctor relationships are strengthened \cite{ridd2009}: (i) continuity of care via consultations with the same doctor \cite{adler2010, lings2003}; and (ii) a sense of trust between patient and doctor (both in the ability to understand the patient's condition and prescribe the appropriate treatment, and in the sense that a patient will adhere to and follow through with that course of treatment). Repeated interactions with the same doctor allow patients to build more secure expectations based on a history of other interactions and anticipation of future interactions \cite{tarrant2010}.

Typically, continuity of care can be measured as the number of successive consultations or the longitudinal duration of the relationship \cite{adler2010, eveleigh2012}. Traditionally, patients' trust in their family doctors has been measured qualitatively through survey data \cite{birkhauer2017}, an approach that suffers from selection bias and self-reporting bias. This poses a challenge for healthcare providers who wish to evaluate the effectiveness of interventions (\emph{e.g.} matching patients with family doctors in our case) using large-scale data analytics. 

Specific factors have been shown to affect patients' trust and confidence in primary care doctors, particularly demographic characteristics and perceived psychosocial factors, such as a sense of being taken seriously, or being involved in decisions regarding their care (as shown by an analysis of over 3 million questionnaires from English national GP patient survey \cite{croker2013}). Furthermore, patients who were given a choice in primary care doctors were more likely to trust them \cite{kao1998}. These findings show that provisioning a ranked list of recommended doctors based on predicted patient-doctor trust would encourage patients to engage with those doctors, leading in turn to more positive clinical encounters and better overall consultation experiences.

\subsection{Recommender Systems in Healthcare}
A recommender system is a class of information filtering system that seeks to predict the fidelity or preference that a user has for an item or entity. It has been widely used to recommend books, videos, or news articles on the Internet (\emph{e.g.} \cite{schafer1999}). In the healthcare domain, applications of recommender systems include assisting the decision-making process in the provision of personalized care \cite{sezgin2013systematic}, identifying key opinion leaders among medical practitioners \cite{guo2016}, supporting patients to find preventative healthcare help in planning personalized therapy \cite{graber2017}, providing personalized healthcare guidance \cite{schafer2017towards} and, more recently, recommending patients with doctors based on their previous consultation history \cite{han2018} .

Broadly, there are three types of recommender systems: Collaborative Filtering (CF), which explores the interaction between patients and doctors, and Content-Based (CB), which explores similarities between entities for which a user expressed a preference in the past \cite{adomavicius2005}. More specifically, CF models analyze the relationships between users and interdependencies among items to identify preference similarity across individuals. Matrix Factorization (MF) is among the most popular realizations of collaborative filtering owing to its scalability and domain-free flexibility. Essentially, MF characterizes both users and items by vectors of latent features, such that a user's interaction with an item is succinctly described by the inner product of their latent vectors \cite{koren2009}. Finally, hybrid approaches combine CB and CF methods to overcome their specific limitations \cite{burke2002}. 

Recommender systems learn about a user's preference among items through \emph{explicit feedback} mechanisms such as ratings or reviews or, in want of these, \emph{implicit feedback} in the form of expressed preferences revealed through behavioral observations \cite{hu2008}. Implicit feedback is more prevalent and easier to collect than explicit feedback, as it does not require explicit information from users beyond their engagement with the system. In our case, as patients' consultations with family doctors in our dataset indirectly reflect patient's opinions, we infer patient-doctor trust from prior interactions, which imply patients' opinions about doctors through their willingness to revisit (or not) a doctor in the future.

However, different recommender system architectures suffer from some notable challenges, such as under performing when operating with sparse data (where there may not be enough information to infer relationships from the utility matrix alone), scaling up in terms of users, items, and their corresponding metadata, and the \textit{cold start} problem of making recommendations to new users who have had no prior interactions with existing items\cite{ghazanfar2010scalable}. Furthermore, and particularly in healthcare, users' privacy must be adequately protected without compromising the quality of the recommendations. In \cite{ramakrishnan2001being}, the authors identify this issue as "the combination of weak ties (an unexpected connection that provides unexpected recommendations) and other data sources that can be used to uncover identities of users in an anonymized dataset." Thus, applications of recommendation systems in areas related to health like medication, treatments, or primary care allocation are still in their infancy concerning trustworthiness and reliability \cite{sezgin2013systematic, schafer2017towards}. From patients' perspectives, such systems should provide explainable recommendations and safeguard against poor recommendations in order to be trustworthy. From the perspective of healthcare professionals, these systems need to provide suitable recommendations based on their domain knowledge and experience. More generally, insurance companies and healthcare institutes are interested in improving recommendation rates through research and reaping the potential benefits of these recommendation systems  \cite{schafer2017towards}.

%% file: 2_data.tex
We have partnered with a leading private healthcare provider and clinical network from Portugal that aims to undertake digital transformation for its health service. The network is made up of 18 hospitals and clinics with over 7,000 staff and serves 2.5 million patients each year. It employs general practitioners in 14 hospitals to perform primary diagnosis and treatment of common illnesses, and who refer complex diagnostic procedures to specialists within the network. Currently, it is planning to initiate an enhanced primary care plan named \emph{``My Doctor''}  to provide each of its patients with a family doctor who will be more readily available to them and who will ideally serve as their long-term healthcare providers.  Essentially, rather than just presenting a directory of family doctors in alphabetical order for patients to choose from, the new primary care plan will empower patients to take an active role in selecting their own family doctors. The matchmaking process requires the healthcare network to learn about the healthcare preferences of each patient and to generate personalized recommendations accordingly.

However, the healthcare network still needs to address several concerns before fully implementing and deploying the matchmaking mechanism as one of the key features in its digital health service. As patients have different levels of engagement with the healthcare network, information availability varies significantly across individuals. Firstly, a majority of patients have never before consulted with family doctors, since primary care is not mandatory before accessing other specialized services. It would be challenging for the network to learn about their preferences without data about past interactions. Secondly, patients who have had previous consultations with family doctors but want to change from their current doctor may be interested in knowing about the preferences of other patients who have visited the same family doctor. For example, patients often find a conflicting schedule with their current family doctor and may benefit from knowing about family doctors that have been visited by other similar patients. Finally, specific groups of patients, such as those with chronic illnesses, require special care and may benefit significantly from personalized primary care \cite{bodenheimer2002}. The healthcare network may consider identifying and matching these patients with family doctors who have previous experience treating other patients with the same chronic conditions.

\subsection{Data Source}
The main data source is an operational database of anonymized transactions between 2012-2017 from each hospital in the network\footnote{Full data from 2012 to 2016 and partial data in 2017 until April.}. Each transaction is defined as an episode that includes the set of services provided to treat a clinical condition or procedure (an \emph{episode of care} is commonly used in the health sector for billing purposes). The healthcare network maintains a decentralized database, with 7 data centers gathering operational transactions from the 18 hospitals (each data center serving between 1-4 hospitals). Patients are assigned a unique ID across the whole network, as well as a local ID at each hospital they visit. We first clean the data by removing out-of-scope episodes (\emph{e.g.} those that either have been canceled or correspond to emergency visits, when patients are simply assigned to the first available doctor and thus are not able to choose their preferred doctor), and deriving consistent patient and doctor IDs across all hospitals. In other words, we identify patients and doctors across appointments at different hospitals. From each episode, we derive one interaction between a patient and a doctor. After data cleaning, we obtain 42 million interactions between around 1.3 million patients and 3,500 doctors. For each patient, we also obtain basic demographic characteristics such as gender, age, municipality of residence, etc. 

\begin{figure*}[!htpb]
  \includegraphics[width=\textwidth]{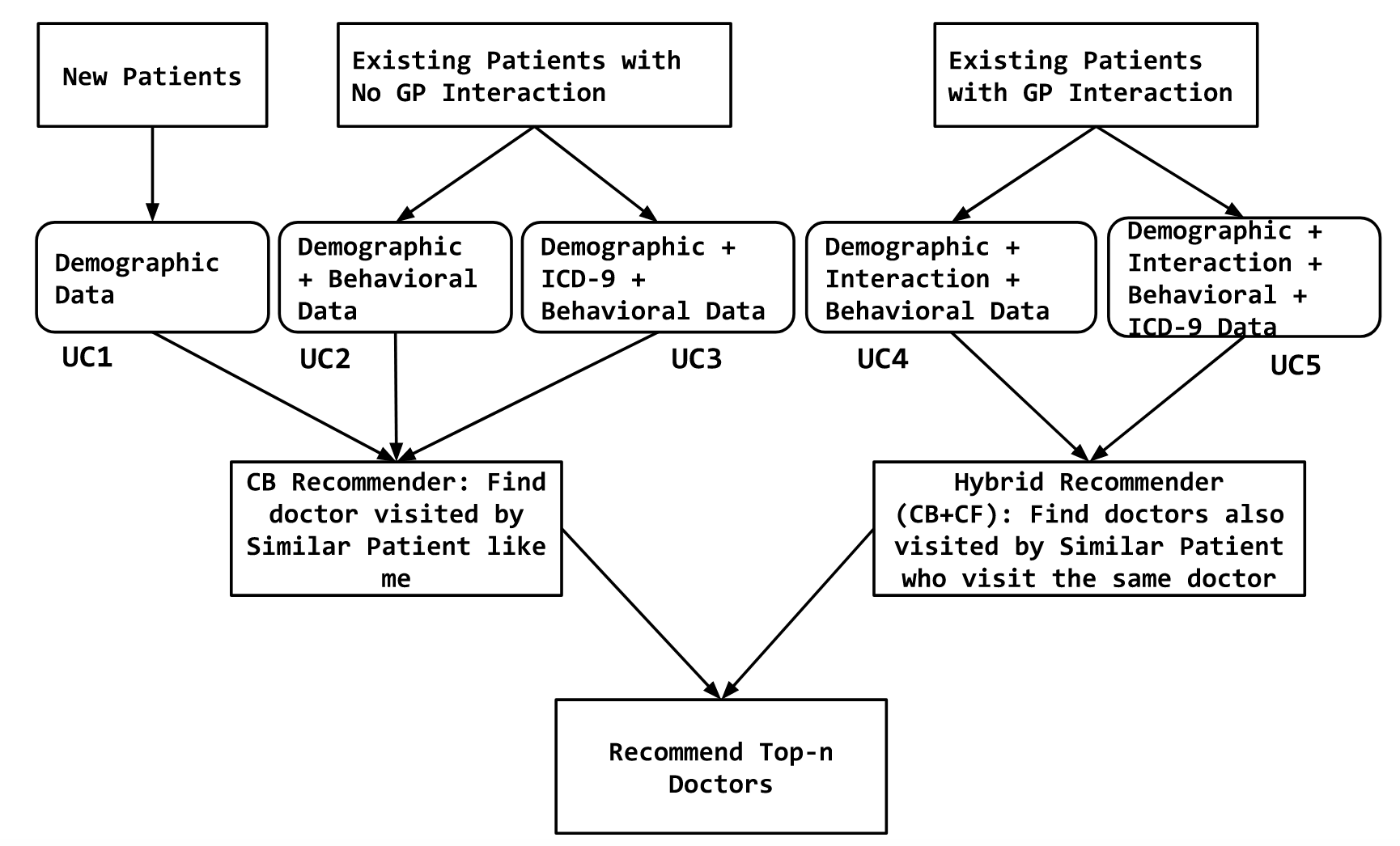}
   \caption{Five use cases to recommend family doctors to patients with different levels of available information, including new patients, existing patients without prior interactions with primary care doctors and existing patients with prior interactions with primary care doctors, separately.}
  \label{fig:usecase}
    \end{figure*}  

Additionally, we obtained doctor registration data from the Human Resources department containing demographic characteristics, education, starting year in the network, as well as their medical specialties. Finally, we identify all episodes performed by primary care doctors (general practitioners, family doctors, and internal medicine specialists).

We further obtained a complementary dataset describing hospital inpatient procedures, which reports International Classification of Diseases (ICD-9) codes. ICD-9 codes allow us to study a subset of patients with certain types of diseases, \emph{e.g.} chronic illnesses. It serves as additional information that we add to the patient profiles to make more relevant recommendations. For example, knowing the medical history of a given patient allows the recommender system to suggest them doctors who are favored by other patients with similar conditions, suggesting proficiency in understanding the implications of those conditions for the patient's primary care needs.

\subsection{Use Cases}
Given the different level of information available to us about different patients, we propose five use cases to make doctor recommendations in different scenarios. Figure \ref{fig:usecase} illustrates each of the use cases that we explain in detail below.

\textbf{Use Case 1 (UC1): New patient} For new patients that have recently joined the network, we have only basic demographic characteristics. We can only find doctors visited by patients with a similar demographic profile, which is essentially a CB recommender.

\textbf{Use Case 2, 3 (UC2, UC3): Existing patient with no interactions with primary care doctors}  For existing patients that have visited specialists but never visited family doctors, we can summarize their past behavior within the network to build a patient profile. For example, we can derive the frequency of visits and hospitals they have visited to help narrow down the list of family doctors from the same hospital. Likewise, the ICD-9 information provides extra information for a complete patient profile. Both UC2 and UC3 can also be considered CB recommenders.

\textbf{Use Case 4, 5 (UC4, UC5): Existing patient with prior interactions with primary care doctors}  For existing patients that have previously visited family doctors, we can leverage interaction data and simply perform CF to find doctors also visited by the patients who visit the same doctor. However, as we also have demographic and behavior data about them as supportive information, we can further combine the advantages of both CB and CF to perform a hybrid recommendation. As such, both UC4 and UC5 follow the hybrid approach that leverages patient and doctor characteristics together with the interactions between them.

%% file: eda.tex
\begin{figure*}[!tbp]
  \centering
  \subfigure[Number of primary care visits by year]{\includegraphics[scale=0.42]{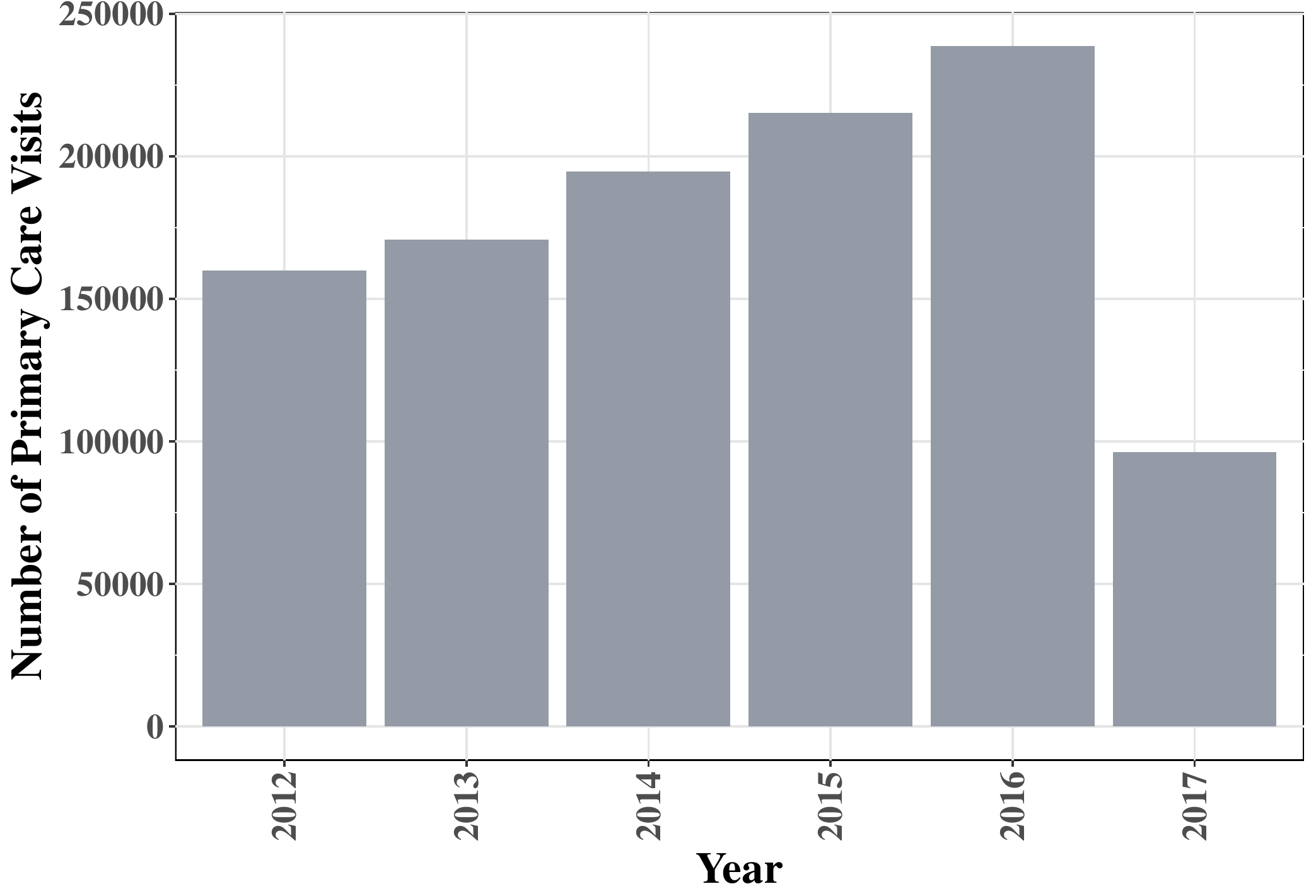}}\quad
  \subfigure[Number of patients and doctors in each hospital]{\includegraphics[scale=0.48]{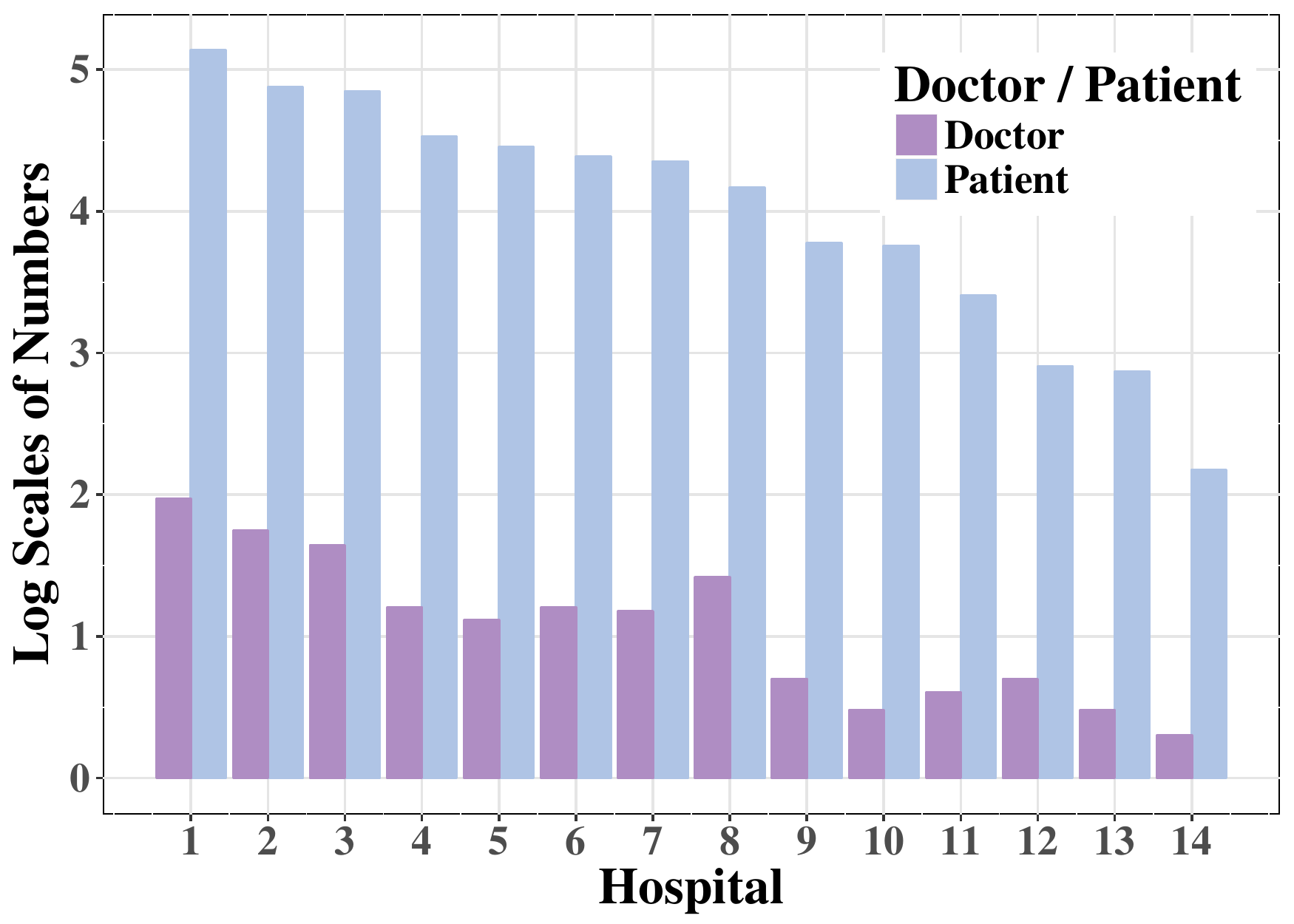}}
  \subfigure[Histogram of patients a family doctor has treated]{\includegraphics[scale=0.42]{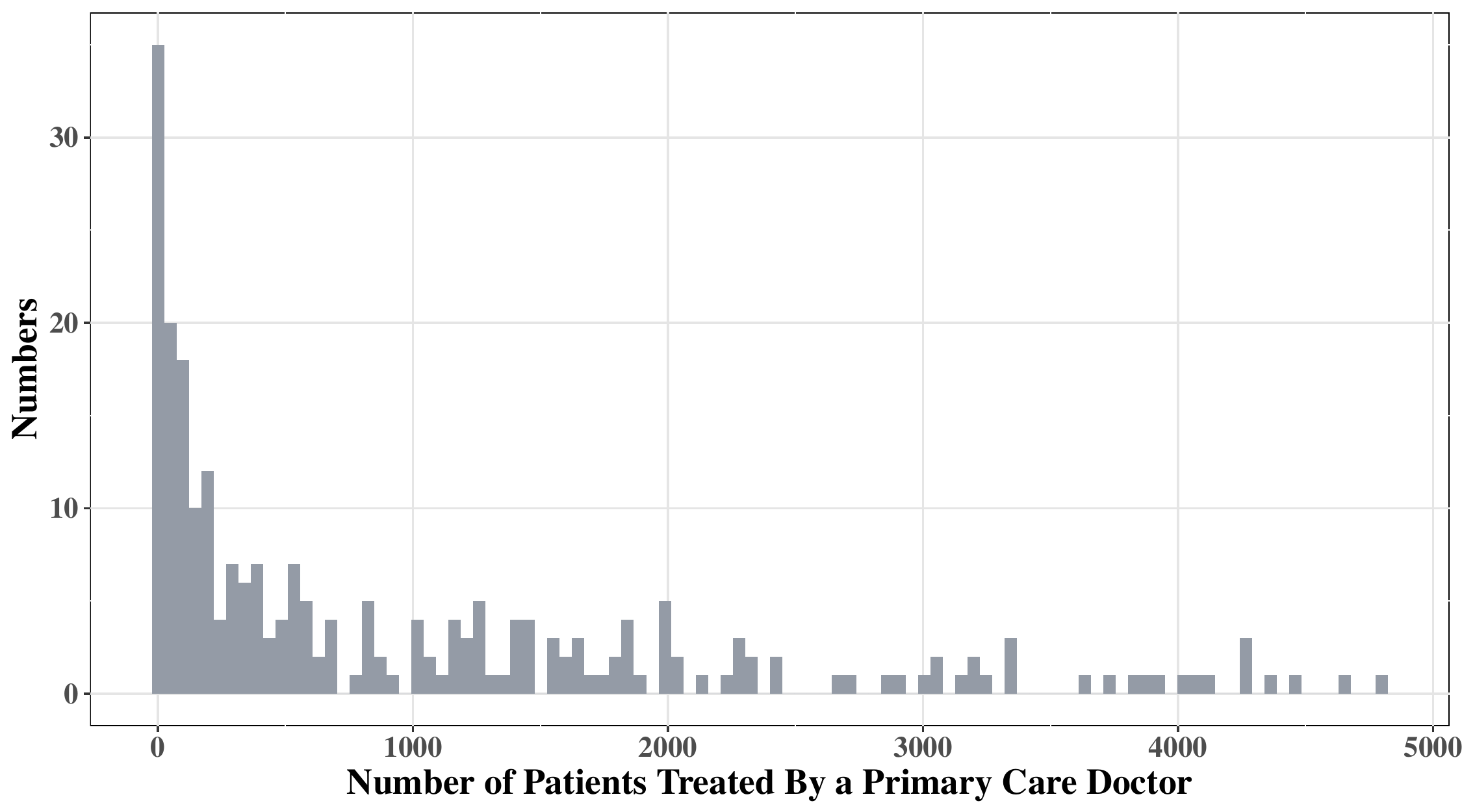}}\quad
  \subfigure[Histogram of number of doctors a patient visits]{\includegraphics[scale=0.40]{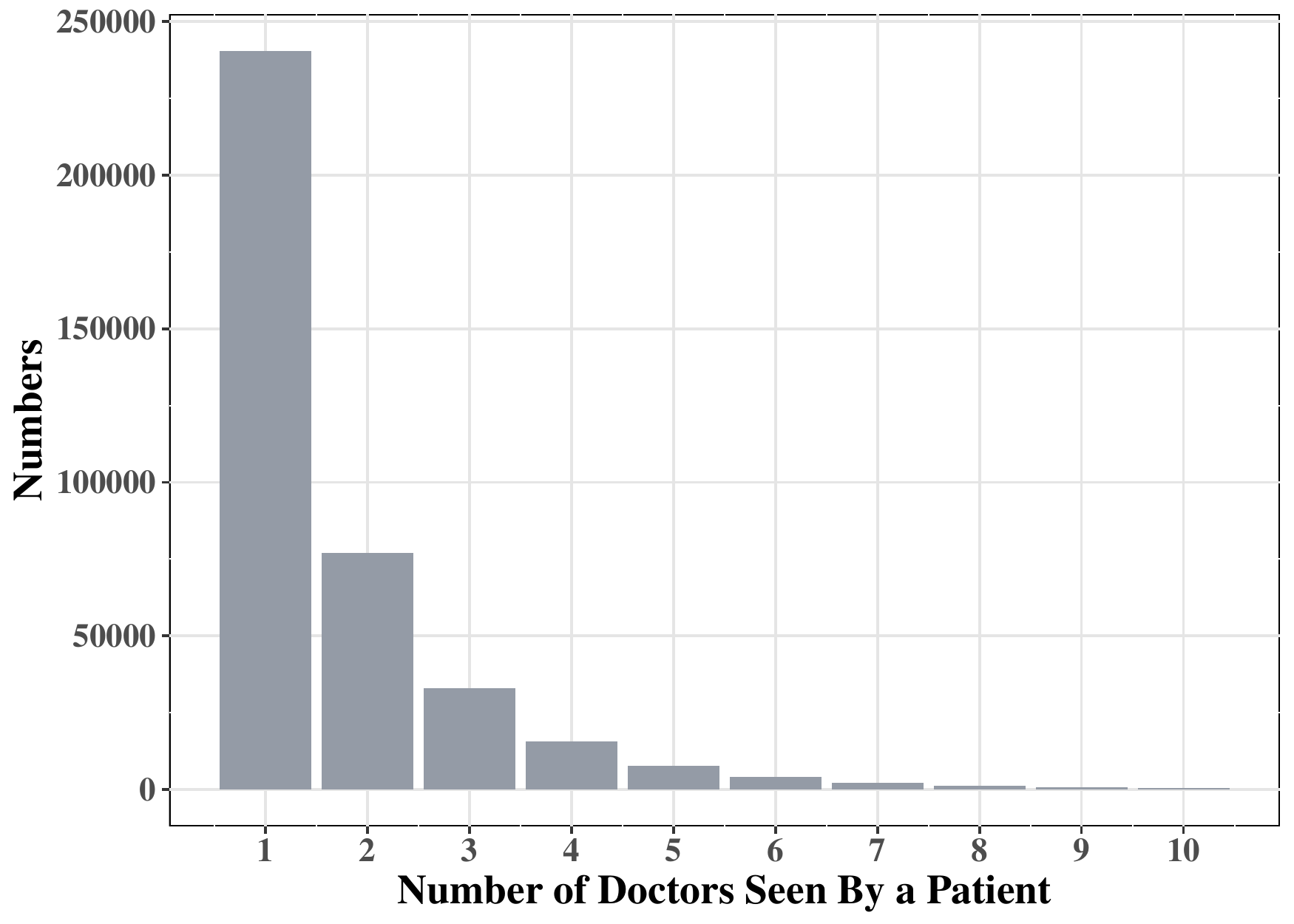}}
 \caption{Descriptive histograms from behavioral characteristics of patients and doctors}
 \label{fig:eda}
\end{figure*}

As we need to evaluate the performance of the proposed matchmaking mechanism, by predicting and validating whether a patient would actually visit the recommended doctor, we retain only the interactions between patients and family doctors as the ground-truth data and the corresponding patient and doctor metadata. This leaves us with 1.07 million consultation records between 382,817 patients and 314 family doctors in 14 hospitals over the years 2012-2017.

For each patient, we have demographic data, such as gender, age, region, as well as past behavioral data, such as the number of visited hospitals, length of time registered with the network, etc. We also append ICD-9 code profiles for a subset of 67,362 patients that we plan to evaluate separately\footnote{In practice, we only use the 24 Major Diagnostic Categories (MDC) codes, given the size of patient data, to ensure we can identify enough similar patients under the same medical condition. However, this does not preclude us from directly using the ICD-9 code if enough patients are admitted as inpatients.}. Moreover, for doctors, we know their demographic data (gender and age), as well as length of time working in the network, number of hospitals they have worked at, number of patients they have treated, and seniority level.

To better understand the demographic and behavioral characteristics of patients and doctors, we next discuss several important insights regarding various features which are significant in the design of the matchmaking system. Figure 2(a) shows the number of visits to family doctors over the observational period. We clearly see a trend of increasing number of primary care visits (with only four months of visits from the year 2017) to the network, which partially explains the motivation for the desire to design a more intelligent matchmaking process in response to increasing demand. Figure 2(b) shows the number of patient and family doctors in each of the 14 hospitals. We find that patients' visits to family doctors are not equally distributed over all hospitals. Instead, the healthcare network tends to assign more family doctors to hospitals located in more populous areas, as there are more patients to be treated. However, we also see an imbalance of patient-doctor ratio in some hospitals. For example, hospitals 4 and 5 suffer from a shortage of family doctors, leading to patients having a more limited range of choices. 

\begin{figure*}[htp]
  \centering
    \begin{tabular}{cc}
    \includegraphics[width=90mm]{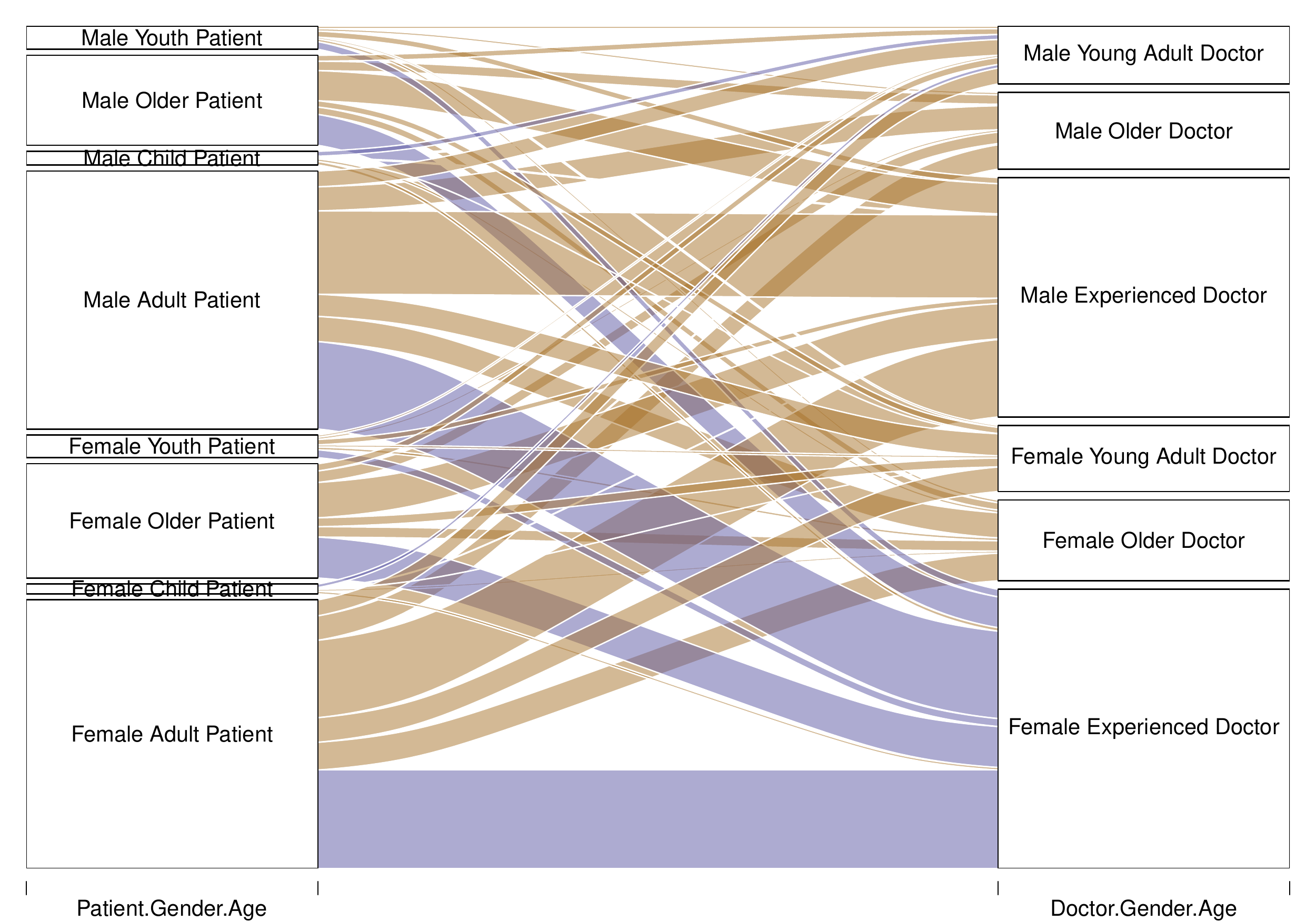}& 
    \includegraphics[width=90mm]{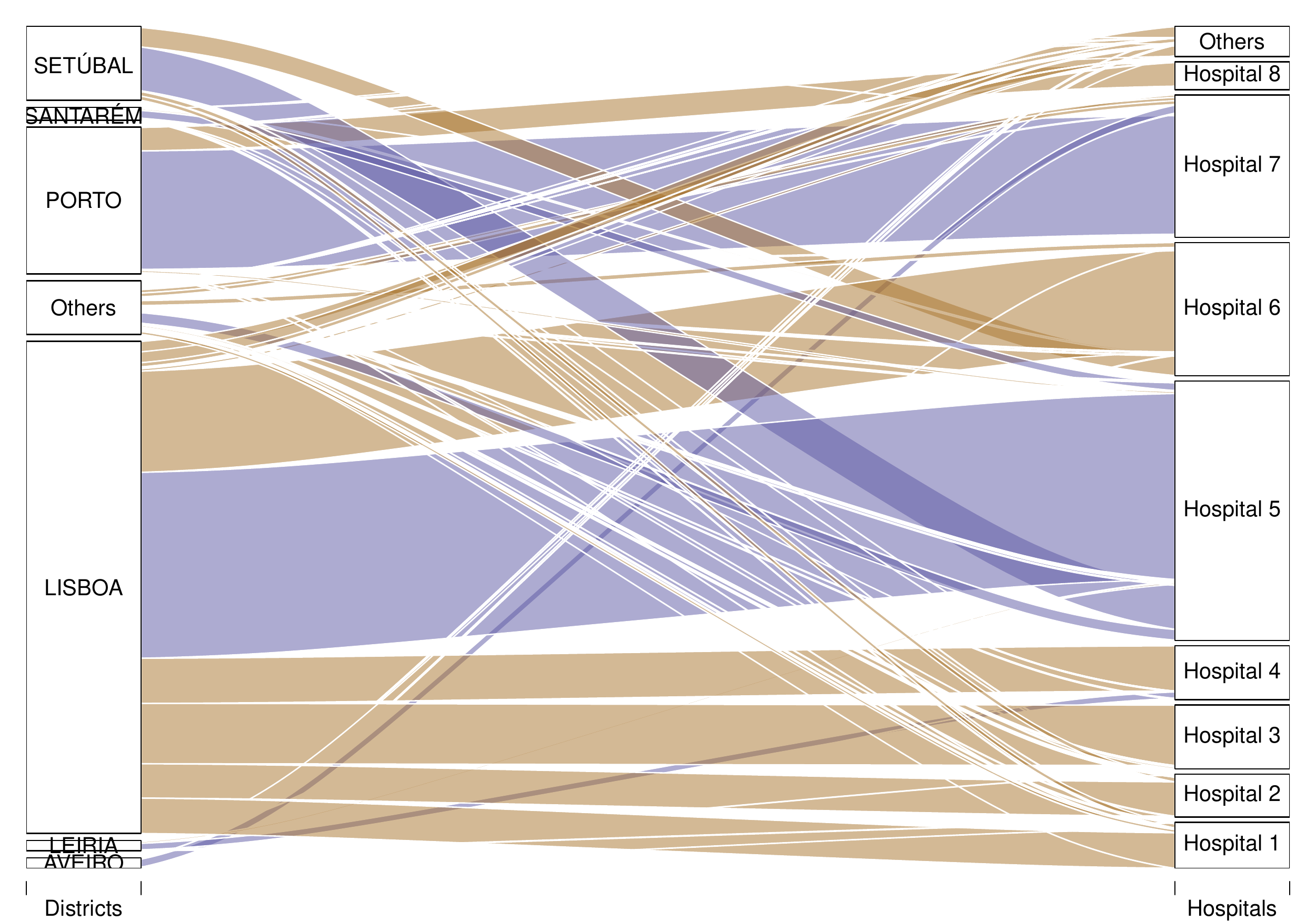}\\
    \end{tabular}
  \caption{Visit patterns of patients to family doctors across gender, age group and home regions}
 \label{fig:eda}
\end{figure*}

Figure 2(c) shows the distribution of number of patients seen by primary care doctors. We see that on average each family doctor has treated over 2,630 patients over the observed period, \emph{i.e.}, about 2 patients per working day (note that doctors work different schedules, some every day, some only once a week). In particular, there is a small group of popular family doctors that treat many patients. Lastly, figure 2(d) shows the number of doctors seen by patients. On average, patients have only visited 1.72 family doctors, 2.81 times each, during the observed period, while nearly half of patients have switched family doctor at least once. This implies that patients do not easily trust one family doctor with just a few consultations. Instead, many of them tend to switch until they find the preferred one.

We further examine the patterns of patient-doctor visits across different demographic groups using a Sankey diagram, where the widths of the bands are directly proportional to total number of patients. On the left panel of Figure \ref{fig:eda}, we show how patients from different gender and age groups choose their family doctors\footnote{We follow  WHO's definition of age group: 0-15 as child; 15-24 as youth; 25-60 as adult and 60 or over as elderly. Doctor's age group is defined as: 25-40: young; 40-60: experienced and 60 or over as senior.}. We find consistent gender homophily for female patients across different age groups, \emph{i.e.,} female patients generally tend to visit female doctors. However, male patients are equally likely to visit both male and female doctors. Also, doctors' age seems to outweigh their gender, such that all patients are more likely to visit doctors whose age is above 40.

Figure \ref{fig:eda} (right panel), shows how patients from different regions choose to visit doctors working at each of the 14 network's hospitals. We find that most patients only visit their local hospital, rather than travel to a different district. However, some patients are willing to follow their primary care doctor to a different geographical area, for example, keeping the same doctor even after moving to live in a different region. These observations show that it is important to take into account the geographical location of patients and doctors, in order to capture these behavioral characteristics. From this, we may infer that most patients would wish to be matched with primary care doctors in nearby hospitals, while some exhibit such a strong preference for their doctor that geography is no longer a deciding factor.

%% file: 3_methods.tex
Given the need to perform the matchmaking across different use cases, it would be inefficient to design separate recommendation algorithms for each of them. Instead, we adopt a unified approach proposed in \cite{kula2015}, to perform hybrid matrix factorization (MF) and recommend each patient a list of family doctors according to the level of information available about them. We achieve this by learning latent representations for patients and doctors from their interactions and metadata. This approach allows us to perform content-based recommendations for patients that have not yet interacted with family doctors. These latent representations are essentially linear combinations of their characteristics. For example, the representation of a 65-year-old female patient is the sum of the representations of \emph{elderly (age$>$60)} and \emph{female}.

Moreover, interaction data further allows us to infer the preferences of one patient as learned from other similar patients. If two doctors are both visited by the same patients, their embeddings are likely to be much closer than doctors who have never been visited by the same patients. In other words, the co-visitation patterns among patients allow us to automate word-of-mouth recommendations.

% \newpage
\subsection{Hybrid Model}
Formally, we denote $M$ and $N$ as the number of patients and doctors, respectively, and the patient-doctor interaction matrix $\vec{Y}\in \mathbb{R}^{M\times N}$ as:
\begin{equation}
y_{ij} =    \begin{cases}
        1, & \text{if interaction (patient $i$, doctor $j$) exists} \\
        0, & \text{otherwise}
        \end{cases}
\end{equation}

We construct feature vectors from patient and doctor metadata. Each patient $i$ is described as a set of features $f_i\subset F^I$, and each doctor $j$ is  given by features $f_j\subset F^J$. For each feature $f$, let $\vec{e_f^i}$ and $\vec{e_f^j}$ denote the $l$-dimensional latent embedding vectors for patient and doctor, respectively. As such, latent representations of patients and doctors can be represented as the linear combination of their latent feature vectors.

Here, we simply let the latent representation of patient $i$ be the sum of its features' latent vectors:
\begin{equation}
\vec{p_i} = \sum_{u\in f_i} \vec{e_u^i}
\end{equation}

Similarly, the latent representation of doctor $j$ is also given by the sum of its features' latent vectors:
\begin{equation}
\vec{q_j} = \sum_{u\in f_j} \vec{e_u^j}
\end{equation}

Then, MF learns $\vec{p_i}$ and $\vec{q_j}$, such that the predicted score for unobserved entries $\hat{y}_{ij}$ is given by the inner product of latent patient and doctor representations:
\begin{equation}
\hat{y}_{ij} = g(i, j | \vec{p_i}, \vec{q_j}) = g(\vec{p_i}\cdot\vec{q_j}) 
\end{equation} 

where $g(\cdot)$ denotes the function that maps model parameters to the predicted score. We choose the sigmoid function as it is suitable for predicting binary data:
\begin{equation}
g(x) = \frac{1}{1+\exp(-x)}
\end{equation}

We then formulate a learning-to-rank task by using Weighted Approximate-Rank Pairwise (WARP) loss \cite{weston2011}. We choose WARP loss over other popular ranking loss, \textit{e.g.}, Bayesian Personalized Ranking (BPR)\cite{rendle2009}, because it directly optimizes the precision@n and is useful when only positive interactions are present (as opposed to \emph{e.g.} negative ratings or below-average ratings after normalization).

More specifically, for each observed interaction $y_{ij}$, WARP samples a negative doctor $d$ and computes the difference between predicted $\hat{y}_{ij}$ and $\hat{y}_{id}$, and performs a gradient update to rank the positive doctor higher if the difference is negative, \textit{i.e.}, a rank violation is found.  Otherwise, it continues sampling negative doctors until it identifies a violating example. Thus, the rank of doctor $j$ for patient $i$ is minimized when taking a large number of sampled doctors $d$ that need to be considered before finding a violating example. 

\subsection{Modeling Patient Trust}
As patients revisit or switch family doctors over the years, their trust in those doctors may be reinforced or diminished accordingly, allowing us to measure trust by accounting for the temporal dynamics of individual patients' consultation histories. For example, a patient would place higher trust in a family doctor that they visit this year than one visited several years ago. Also, patients who repeatedly visit the same family doctor multiple times would develop greater trust in them than in one they visited just once. Therefore, we can model the trust $\text{T}_{ij}(t)$ between a patient $i$ and a family doctor $j$ at time $t$, given both the frequency and recency of their consultation history as:

\begin{equation}
  \text{T}_{ij}(t) = \sum_{t}\sum_k\frac{C_{ij}(t) \mathrm{e}^{-\lambda t}}{C_{ik}(t)}
\end{equation}

where $\lambda$ is annualized discount rate for the exponential decay function and treated as hyper-parameter during the model training, $C_{ij}(t)$ is the number of consultations between patient $i$ and doctor $j$ until year t, which is normalized by the total number of her consultations with $k$ doctors $C_{ik}(t)$ thus far. We choose the exponential decay by assuming that the strength of a social relationship decreases exponentially as time increases \cite{centola2010}.

The quantitative trust measure is introduced as a weighting on individual interactions in the interaction matrix. This is similar to the instance weighting suggested in \cite{koren2009b}， with the weight being learned automatically from the data. When we train the model using stochastic gradient descent with weighted sampling, the trust measure may bias the learning rate toward highly trusting patients and thus improve the convergence rate \cite{needell2014}.

\subsection{Evaluation}
\begin{figure}[!htpb]
  \includegraphics[width=0.55\textwidth]{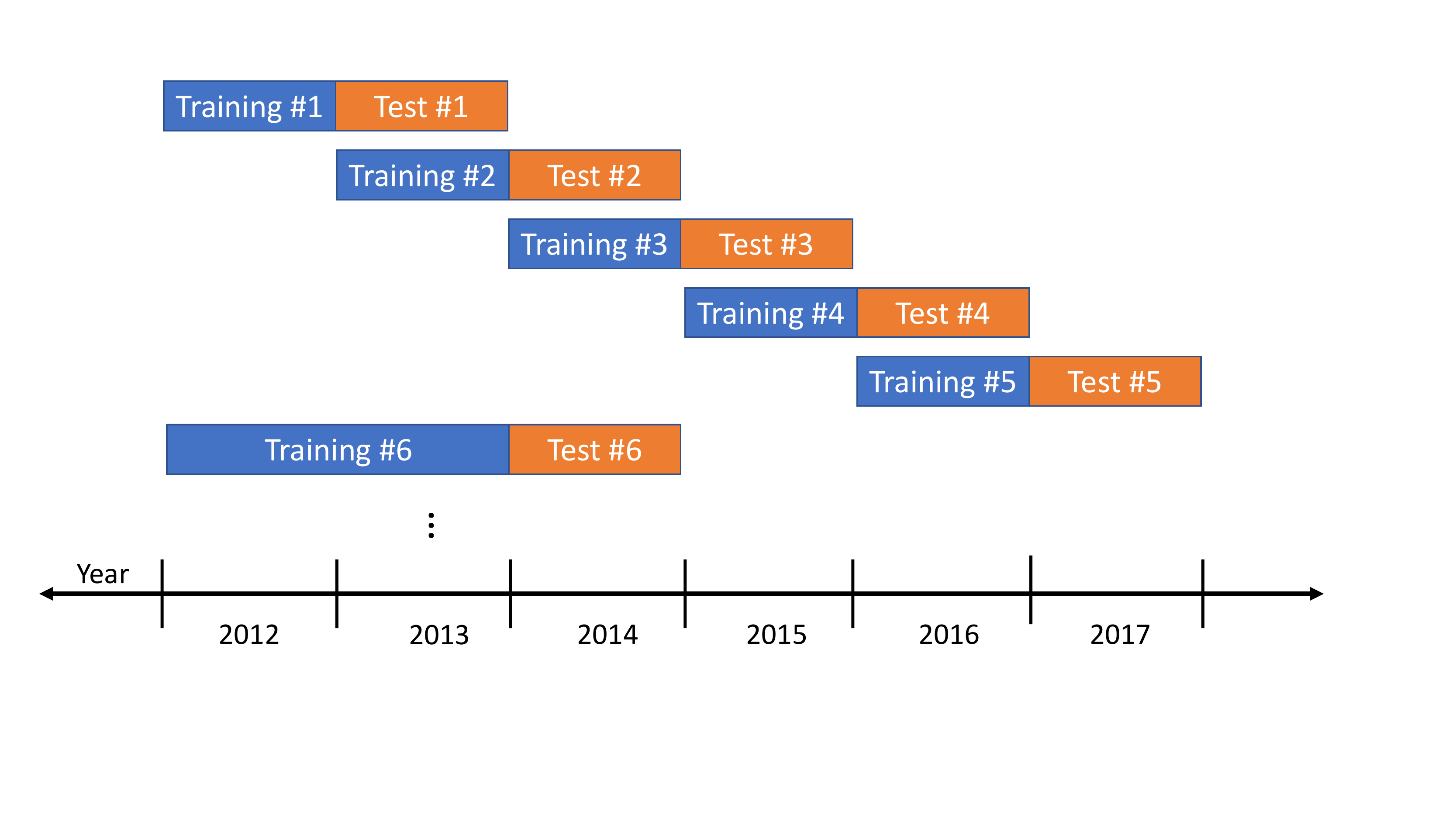}
   \caption{Temporal cross-validation by dynamically and chronologically splitting dataset into training and test set over the observational period. }
  \label{fig:tscv}
    \end{figure}

To determine hyper-parameters of our model, we perform temporal cross-validation by chronologically splitting the data into train and test sets over the years, as illustrated in figure \ref{fig:tscv}. More specifically, for each patient-doctor triplet $<i,j,t>$ in the training set (\textit{e.g.,} $t$ represents year 2014), we retain the corresponding triplet $<i,j,t+1>$ as the test set (\textit{e.g.,} $t+1$ represents year 2015). The walk-forward optimization strategy allows us to tune the hyper-parameters while avoiding information leakage because the test set is ensured to always be preceded by a training set observed in the past.

The model was trained on an AWS EC2 c4.xlarge instance, and we obtain the following hyper-parameters through grid search for the model performance evaluation in the next section: 
\emph{\{learning\_rate: 0.012, epoch: 120, $\lambda$: 0.3, no\_components: 95, max\_sampled: 3\}} (with per-parameter adaptive learning rate schedule as ADAGRAD \cite{duchi2011}), where \emph{no\_components} is the dimensionality of the feature latent embeddings and \emph{max\_sampled} is the maximum number of negative samples used during WARP fitting.

%% file: 4_results.tex
\begin{figure*}[!htbp]
  \centering
 \includegraphics[scale=0.49]{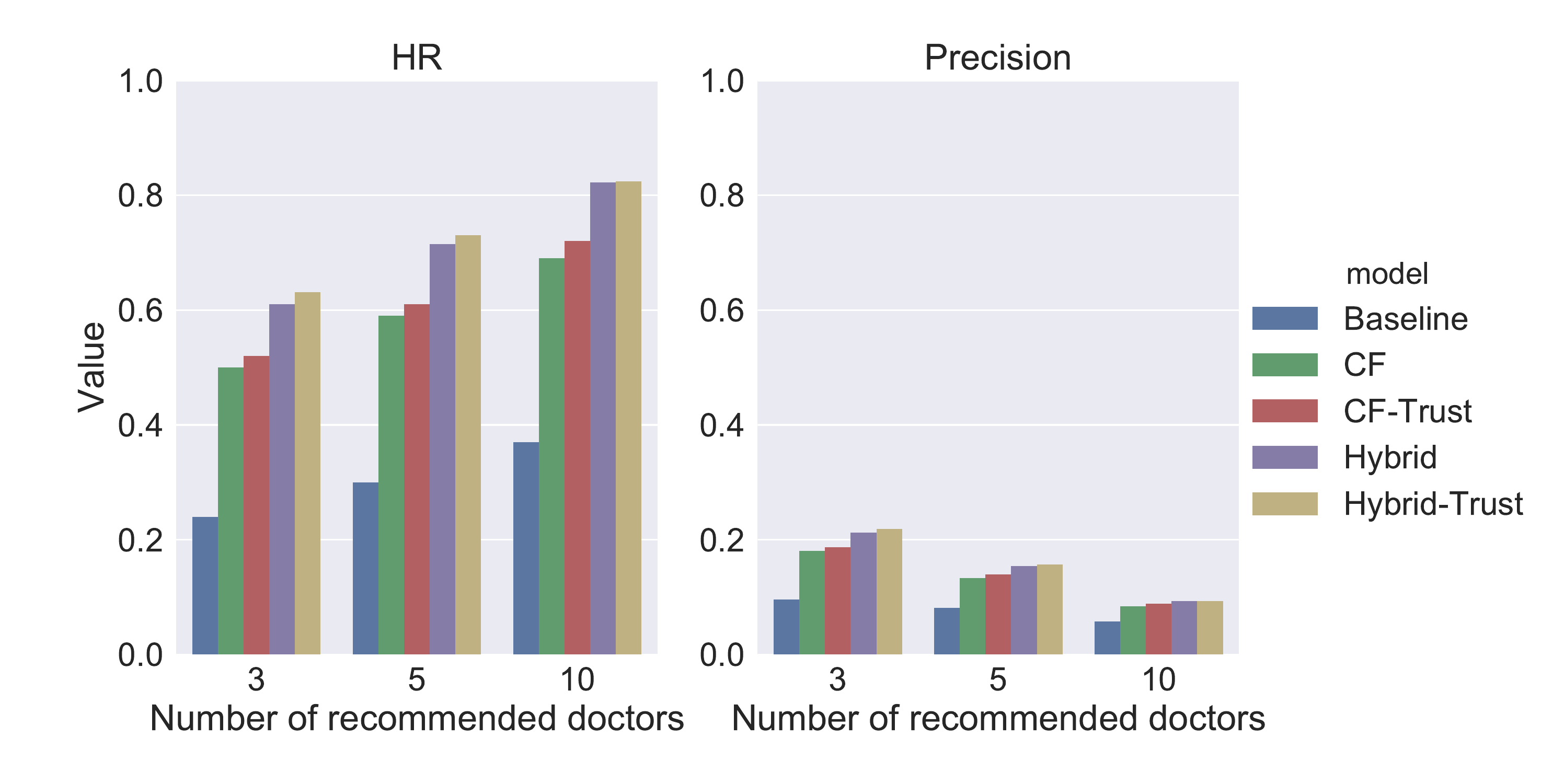}
 \label{fig:result}
 \caption{Evaluation of Top-N family doctor recommendations for performance comparison (Hit Rate on the left and Precision on the right)}
\end{figure*}

We compare the performance of our systems across the following models:
\begin{itemize}
\item \textbf{Baseline} Although a non-personalized popularity-based benchmark has been commonly used in the existing literature, we argue that it is not suitable in our case because doctors typically have limited availability for appointments in practice. Instead, we construct a baseline model with a set of ordered heuristics, which are iterated until a desired number of recommended doctors are obtained: 
\begin{enumerate}
\item recommend the most frequently visited doctors; 
\item if multiple doctors have been visited the same number of times, recommend the most recently visited among them; 
\item if multiple family doctors have been visited equally frequently and recently, randomly select among them; 
\item if patients have visited less than n doctors, append the most popular doctors to the list until enough recommendations are generated.
\end{enumerate}
\item \textbf{CF}: standard collaborative filtering model derived from the patient-doctor interaction matrix. 
\item \textbf{CF-trust}: CF model with the elements of the interaction matrix replaced with the proposed trust measure.
\item \textbf{Hybrid}: a hybrid approach that uses both patient and doctor interactions as well as their metadata.
\item \textbf{Hybrid-trust}: Hybrid with the elements of the interaction matrix replaced with the proposed trust measure. 
\end{itemize}

We generate a list of doctors for each patient ranked by the predicted trust and evaluate the performance by Hit Rate@n (HR@n) and precision@n (p@n) with $n\in\{3,5,10\}$. HR@n refers to the ratio of correct predictions where patients visit any family doctor from the $n$ recommendations. p@n refers to the number of correctly predicted family doctors that patients visit. Both metrics intuitively measure whether the doctors that are recommended (after observing the training set) are actually visited by a patient (as evidenced by the testing set).

Figure 5 compares the performance of the five models in terms of Hit Rate (left) and Precision (right). We find that both CF and hybrid models consistently outperform the heuristic baseline model by significant margins. This implies that collaborative information from other similar patients together with patient and doctor characteristics may well complement patients' own consultation histories to generate more accurate doctor recommendations. Moreover, introducing the quantitative trust measure further improves model performance by recommending 2-3\% more patients with relevant doctors. This indicates that modeling patient trust may contribute to understanding the temporal dynamics of patient-doctor relationships.

Furthermore, we separately evaluate the model performance on the subset of patients with ICD-9 codes (18\% of total) to understand the impact of possessing data about patients' medical history on the ability to make informed recommendations. Figure 6 shows the performance comparison across the five models for these patients. As expected, we have very similar findings but with overall better hit rate and precision compared to evaluating on the full dataset (Figure 5). This is likely because these patients who have been admitted as inpatients tend to have more health records and may visit family doctors more often. However, we find that the hybrid model performs even better than the CF model using ICD-9 code. This is particularly interesting because we show that feature engineering with the domain knowledge from the health sector may help boost the model performance. 

\begin{figure*}[!htbp]
  \centering
 \includegraphics[scale=0.49]{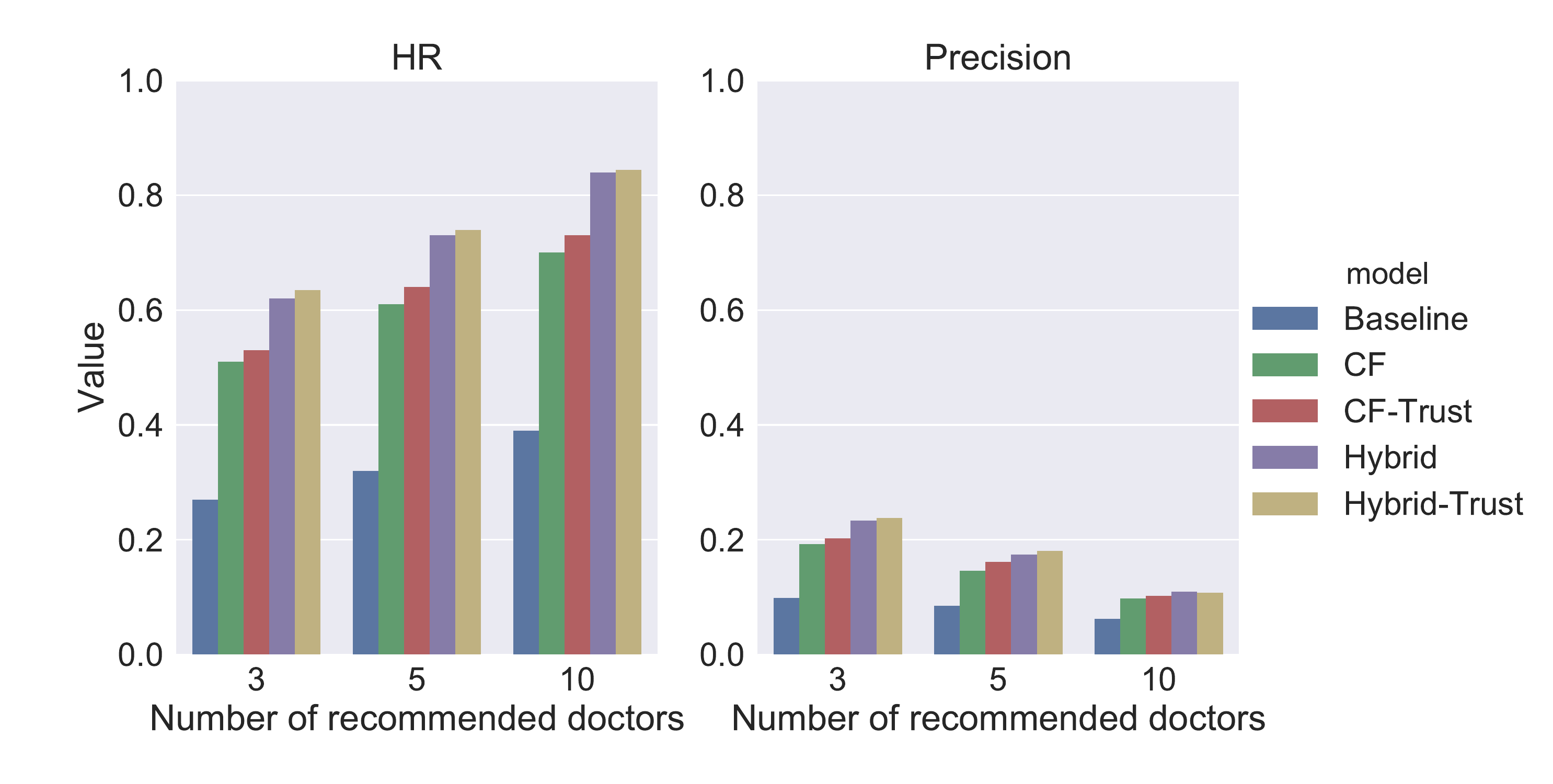}
 \label{fig:result2}
 \caption{Evaluation of Top-N family doctor recommendations for performance comparison for subset of patients with ICD-9 code information (Hit Rate on the left and Precision on the right)}
\end{figure*}

%% file: conclusions.tex
Although person-centered primary care is becoming increasingly important for promoting universal healthcare and improving health outcomes, even health systems from high-income countries demonstrate considerable gaps between the actual status of primary care and the WHO vision \cite{gauld2012}. As such, it remains a challenge for healthcare providers to transform the service provision into a more person-centered primary care, in particular with the aim of promoting the development of long-lasting patient-doctor relationships.

Greater trust in a patient-doctor relationship correlates with perceived quality and continuity of care \cite{hall2002b}. The underlying logic is simple: patients who trust their doctors are more likely to follow their advice and develop long-lasting relationships with them. There is also a relationship-reinforcing-effect that needs consideration. Continuity of care and familiarity helps doctors better understand their patients' needs and helps patients act preventatively and live healthier lives, thus reinforcing the strength of the relationship. Today, many developing countries are facing aging populations, and healthcare costs are expected to increase in order to meet their needs. Hence, investing in scalable systems with the potential to improve patient-doctor relationships and strengthen the gate-keeping role designated to primary care doctors seems very much needed. We see this work as a step towards building such a system.

In this work, we partnered with a private healthcare provider in Portugal to design a matchmaking mechanism between patients and primary care doctors in order to promote continuity of care. More specifically, we describe the matchmaking process as five distinct use cases adjusted to the different levels of information that may be available about a patient. To this end, we adopt a hybrid approach that aims to provide a unified solution that presents all patients with a list of personalized doctor recommendations. More importantly, we further model patients' trust in their doctors by using a large-scale dataset of consultation histories, while accounting for the temporal dynamics of their relationships. This allows us to quantitatively measure trust by examining the frequency and recency of patient's interactions with doctors, thus moving beyond existing qualitative findings which have traditionally relied survey data to gauge patient-doctor relationships.

We gathered multiple data sources into a dataset that contains patient and doctor characteristics, as well as the interactions between them. The interaction data can be used to infer patients' preferences toward primary care doctors through collaborative information from patients that have also visited the same doctor. Meanwhile, the patient and doctor characteristics can serve as additional information to find similar patients or doctors across social dimensions. We identify some key characteristics that may correlate with patients' preferences from domain experts and exploratory data analysis. In other words, through feature engineering, we obtain a set of useful features to best represent the consultation history records for the hybrid recommender system. 

Our results show superior predictive accuracy compared to both the heuristic baseline and a classical CF recommender system. This also holds when we separately examine a subset of patients with auxiliary information on their medical history. More interestingly, the proposed quantitative trust measure further improves model performance. Consequently, we are able to recommend over 80\% of patients with relevant primary care doctors by presenting them with a list of 10 recommendations (compared to just 37\% using the heuristic baseline or 69\% using CF without the trust measure).

In future work, we plan to deploy the matchmaking service as a key feature into the production environment of digital health services at network to gather patients' expressed preferences and evaluate the utility of recommendations in randomized controlled trials and eventual deployment. This would not only help to overcome the limitations of offline evaluation of the model performance, but also provide us with patients' explicit preferences to combine with the implicit feedback\cite{jannach2016} and thus help further personalize the model to reflect patients' preferences.